\documentclass[aps, prl, twocolumn, floatfix, superscriptaddress, showpacs]{revtex4-1}

\usepackage{graphicx}
\usepackage{bm}
\usepackage{enumerate}
\usepackage[latin1]{inputenc}
\usepackage{color}
\usepackage[colorlinks=false]{hyperref}

\usepackage{amsmath,amssymb}

\begin{document}


\title{Interplay of Electron and Nuclear Spin Noise in GaAs}

\author{Fabian Berski}
\affiliation{Institut f{\"u}r Festk\"orperphysik, Leibniz Universit\"at Hannover, Appelstr.~2, D-30167 Hannover, Germany}

\author{Jens H\"ubner}
\email{jhuebner@nano.uni-hannover.de}
\affiliation{Institut f{\"u}r Festk\"orperphysik, Leibniz Universit\"at Hannover, Appelstr.~2, D-30167 Hannover, Germany}

\author{Michael Oestreich}
\email{oest@nano.uni-hannover.de}
\affiliation{Institut f{\"u}r Festk\"orperphysik, Leibniz Universit\"at Hannover, Appelstr.~2, D-30167 Hannover, Germany}

\author{Arne Ludwig}
\author{A. D. Wieck}
\affiliation{Angewandte Festk\"orperphysik, Ruhr-Universit\"at Bochum, 44780 Bochum, Germany}

\author{Mikhail Glazov}
\email{glazov@coherent.ioffe.ru}
\affiliation{Ioffe Institute, Polytechnicheskaya~26, 194021 St.-Petersburg, Russia}
\affiliation{Spin Optics Laboratory, St. Petersburg State University, Ul'anovskaya~1, 198504 Peterhof, St.-Petersburg, Russia}


\begin{abstract}
We present spin noise (SN) measurements on an ensemble of donor-bound electrons in ultrapure GaAs:Si covering temporal dynamics over six orders of magnitude
from milliseconds to nanoseconds. The SN spectra detected at the donor-bound exciton transition show the multifaceted
 dynamical regime of the ubiquitous mutual electron and nuclear spin interaction typical for III-V based semiconductor systems. The experiment distinctly reveals the finite Overhauser-shift of an electron spin precession 
at zero external magnetic field and a second contribution around zero frequency stemming  from the electron spin components parallel to the nuclear spin fluctuations. Moreover, at very low frequencies features related with time-dependent nuclear spin fluctuations are clearly resolved making it possible to study the intricate nuclear spin dynamics at zero and low magnetic fields. The findings are in agreement with the developed model of electron and nuclear SN.
\end{abstract}
\maketitle

Harnessing coherence is one of the central topics in current research and attracts high interest due to the complex fundamental physics bridging quantum mechanics and statistics as well as due to prospective applications for information processing~\cite{awschalom_semiconductor_2002, devoret_superconducting_2013, leibfried_quantum_2003}. The solid state quantum states based upon the spin degree of freedom of confined carriers in semiconductors are at the forefront of many current research activities in this field. In this respect, optically addressable electron and hole spin quantum states in III-V based semiconductor systems bear the beauty of  efficient options for initialization, manipulation, and readout by light in combination with exceptional sample quality~\cite{RevModPhys.85.79}.  Currently, a promising system for these tasks are donor-bound electrons in ultrahigh quality, very weakly $n$-doped GaAs since the widely spaced, quasi-isolated electrons act as an  ensemble of identical, individually localized atoms~\cite{PhysRevB.82.121308, fu_ultrafast_2008}. However, the ostensible catch of this approach is the inherent interaction with the nuclear spin bath which has been addressed in many different systems so far~\cite{merkulov_electron_2002, PhysRevLett.88.186802, PhysRevB.88.085323,PhysRevLett.94.116601,PhysRevB.84.033302}.

In principle, there are different approaches to deal with the decoherence imposed via the hyperfine interaction. On the first sight, the most obvious way is to replace the isotopes carrying a nuclear spin with spinless isotopes like in $^{28}$Si \cite{morley_initialization_2010} but silicon has the drawback of an indirect gap. Direct semiconductors with spinless isotopes like, e.g., isotopically purified II-VI systems have yet the drawback of inferior sample quality. In single III-V based quantum dots, the hyperfine interaction can be reduced by either moving on to hole spins which show a diminished hyperfine interaction \cite{Brunner03072009, Chekhovich:2013ys,Dahbashi.PRL.2014} or by polarizing the nuclei in order to make them less effective \cite{Sallen.NatComm.2014, PhysRevB.91.205301}. Besides that, the mutual interaction between carrier and nuclear spins is also strain dependent and strongly varying coupling strengths in such nanostructures result in a row of widely discussed problems with the central spin problem being one of the most prominent and complex examples~\cite{Sinitsyn.PRL.2012, PhysRevB.88.085323}. By contrast, donor-bound electrons in high purity bulk GaAs have
 an isotropic, well defined hyperfine interaction in a strain free environment, in which case an in-depth understanding and exploitation of the generic electron and nuclear spin dynamics looks feasible.

Measurements of the intrinsic spin dynamics of weakly interacting donor-bound electrons in bulk GaAs are extremely challenging since any optical excitation of free electrons or holes dramatically affects their spin dynamics. One reason is the slow cooling time of free carriers at low lattice temperatures. The Hanle effect~\cite{PhysRevB.66.245204} and the resonant spin amplification technique~\cite{Kikkawa98} yield long spin relaxation times of resident electrons, however, these experiments involve considerable optical excitation of the sample.  Here, we avoid the problem of optical excitation of free carriers by utilizing spin noise spectroscopy. This quantum optical method \cite{aleksandrov_magnetic_1981, oestreich_spin_2005,PhysRevB.79.035208, hubner_rise_2014, li_intrinsic_2012} based on spectral analysis of fluctuations in the polarization state of the laser light transmitted through the sample is a matchless tool for this task since the technique measures the spin dynamics at thermal equilibrium practically perturbation-less, and provides a complete picture of the involved spin dynamics of both the electrons and the host lattice nuclei on time scales spanning from nanoseconds to milliseconds, respectively.

\begin{figure}[h]
\centering
 \includegraphics[width=0.9\columnwidth]{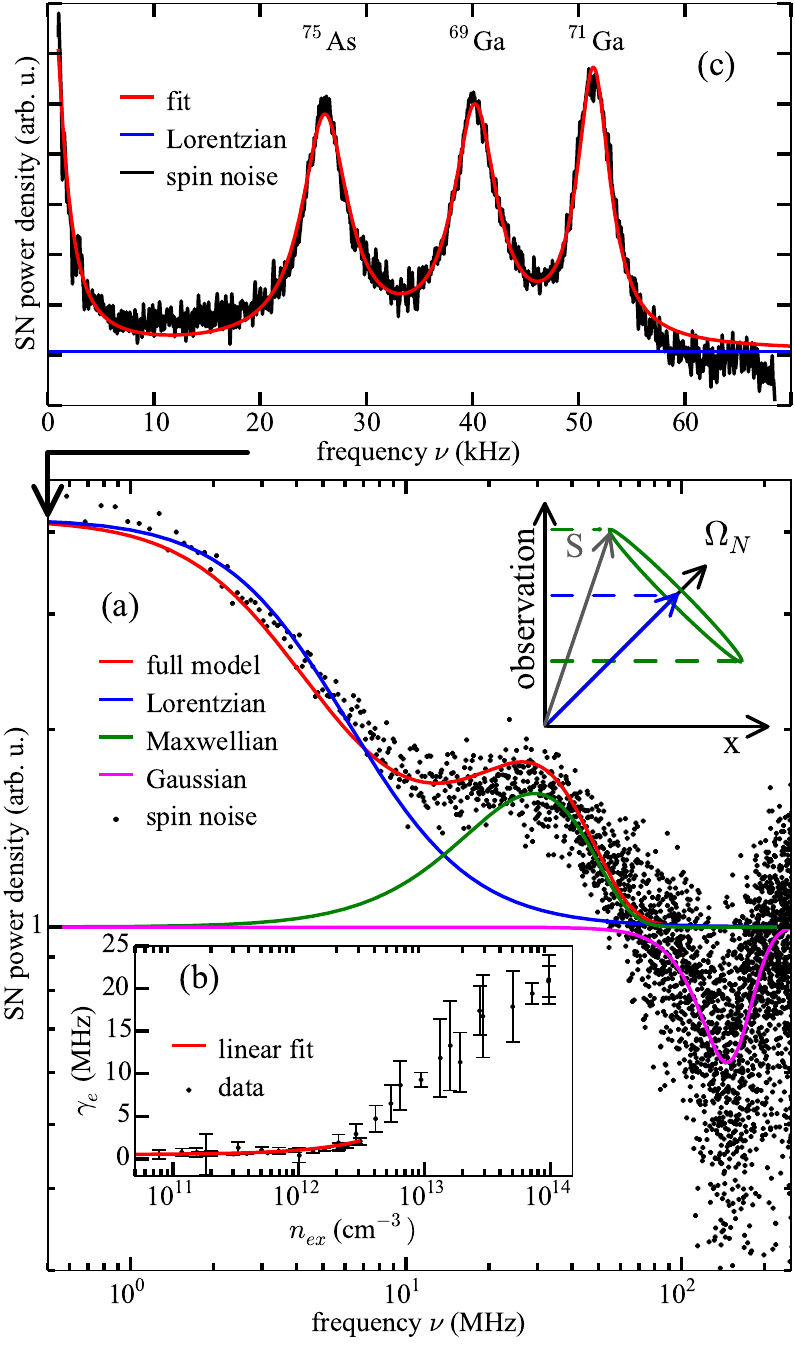}
  \caption{(Color online)
(a): Black dots show SN difference spectrum of donor-bound electrons (SN at $B$=0~mT minus SN at $B$=22~mT) measured at $P=1~\mu$W. The red line is a  fit according to the model of Ref.~\cite{glazov_spin_2015}; blue, green, and magenta curves represent the individual components (see text for details). The difference spectrum is rescaled and shifted to positive noise power densities.
(b): Measured width of the homogeneous SN contribution [blue curve in (a)] as a function of the photo-generated exciton density $n_{\rm ex}$. The red line is a linear fit extrapolating to $\gamma_0=0.50\pm 0.09~$MHz for $n_{\rm ex} \to 0$.
 (c): SN difference spectrum (black line)  in the low-frequency range measured at $P=1~\mu$W and $B=3.75~$mT. Contributions from the individual host lattice isotopes are clearly resolved and labeled respectively. See text for details on the $\nu=0$ feature. All measurements were performed at $T=4.2$~K and quasi-resonant probe. }
\label{fig:spectrum}
\end{figure}

The sample under study is a $10~\mu$m thick, very high purity GaAs layer grown by molecular beam epitaxy (MBE) on top of a semi-insulating GaAs substrate separated by a GaAs/AlAs superlattice and an AlAs etch stop buffer layer. The intentional n-type doping density of the GaAs:Si layer is $n_{\rm D} \approx 1 \times 10^{14}$~cm$^{-3}$ which yields an average distance between two neighboring Si-donors of about 20 Bohr-radii.
The hydrogen-like wavefunction of each localized electron overlaps with $\sim10^{5}$ host lattice nuclear magnetic moments leading to the contact hyperfine interaction~\cite{abragam_principles_2006}.
A via-hole with diameter $d\approx 100~\mu$m is wet chemical etched through the backside of the sample~\cite{etch}  to gain unobstructed optical access to the MBE-grown GaAs for transmission measurements.

The sample is mounted in a cold finger cryostat and cooled down to temperatures between $3.8~$K and $11~$K. An electro-magnet is used to apply transverse magnetic fields $B$ up to $40~$mT with respect to the direction of light propagation. Linearly polarized laser light is focused to a beam-waist of $4.5$~$\mu$m and tuned quasi--resonantly to the donor-bound exciton ($D^0X$) transition at $E_0=1514.26$~meV showing an inhomogeneous  broadening with a full width at half maximum (FWHM) of $\Delta_{\rm FWHM}\approx 150~\mu$eV measured by absorption spectroscopy. The spin induced stochastic Faraday rotation (spin noise) of the transmitted laser light is resolved by a polarization bridge and a low-noise balanced photo receiver. The noise background due to optical and electronic noise is eliminated by subtracting SN spectra with different transverse magnetic field from each other~\cite{muller_semiconductor_2010, suppl}.

Figure~\ref{fig:spectrum} shows the measured spin noise spectra in the frequency range from 1~kHz to 250~MHz. First, we focus on the SN power density between 0.6~MHz and 250~MHz (Fig.~\ref{fig:spectrum}a, black dots) which clearly consists of three contributions. The depicted spectrum is the difference of SN spectra acquired at $B=0~$mT and $B=22~$mT, respectively \footnote{The difference spectrum is normalized to the maximum SN density and shifted by 1 for presentation on logarithmic scale.}. First, we focus on the contributions measured at zero external magnetic field which are centered at $\nu=0$ and at $\nu=\nu_O \approx 30$~MHz. The peak at $\nu_O$ arises from electron spin precession in the quasi-static field of nuclear fluctuations, i.e. from the precession in the randomly distributed Overhauser fields $\Omega_N$~\cite{merkulov_electron_2002} (see pictogram in Fig.~\ref{fig:spectrum}): Their random distribution at different donor sites results in a Maxwell-like SN shape ~\cite{glazov_spin_2012,glazov_spin_2015} $S_M(\nu) =A_M  {(\pi^{2/3}\delta_e/2)^{-3}}  {\nu}^{2}  {\exp({- \nu^2/\delta_e^2}})$ (Fig.~\ref{fig:spectrum}a, green line). Here, $A_M$ is the respective noise power~\cite{normA}, and $\delta_e$ results from the dispersion of the nuclear fields $\Delta_B$ as $ \delta_e = g_e \mu_B \Delta_B/(2\pi\hbar) = 1/(2\pi T_\Delta)$, where $g_e$ is the electron $g$-factor. The $\nu=0$ feature arises from the electron spin component parallel to $\Omega_N$, which is conserved in the course of fast electron spin precession (pictogram in Fig.~\ref{fig:spectrum})~\cite{merkulov_electron_2002,glazov_spin_2012}. This contribution is approximated by a Lorentzian shape (blue line in Fig.~\ref{fig:spectrum}a)\, i.e., $S_L(\nu) = A_L 2{\gamma_e}/{[\pi(\nu^2+\gamma_e^2)]}$~\cite{normA} with $2\pi\gamma_e$ being the damping related to a finite electron correlation at a given donor~\cite{glazov_spin_2015} and to spin-flip processes not related to hyperfine interaction~\cite{glazov_spin_2012}. Such a specific two-peak structure is a distinct feature of the SN of localized electrons coupled to lattice moments and, moreover, the simultaneous presence of both contributions in the zero field spectrum clearly demonstrates that the correlation time is long, i.e., $\tau_c \gg T_\Delta$~\cite{glazov_spin_2015}.

The feature at about $140~$MHz results from the Larmor precession of the stochastically oriented electron spin ensemble in the effective magnetic field given by the sum of the local hyperfine fields and the applied transverse magnetic field. In agreement with Refs.~\cite{glazov_spin_2012, glazov_spin_2015} this feature is well approximated by a Gaussian function ${S_G} (\nu)=A_G/\sqrt{2 \pi\delta^*_e} \times \exp{[ - 1/2~ {(\nu-\nu_L)^2}/{{\delta^*_e}^2}]}$ shown as magenta line in Fig.~\ref{fig:spectrum}a. Here, $A_G$ is the SN power of the precession contribution, $\nu_L=g_e \mu_B B/2\pi \hbar$ is the Larmor-frequency, and $\delta^*_e$ is the spread of the spin precession frequencies caused by nuclear fields and $g$-factor variations, see below.~\cite{glazov_spin_2012} By fitting $S_G(\nu)$ to the data, an electron $g$-factor of $ \left|g^*_e \right|=0.46 \pm0.064$~\cite{errorB} is extracted in agreement with expectations for donor-bound electrons in GaAs.~\cite{PhysRevLett.95.187405}

\begin{table}
\caption{Fit parameters extracted from Fig.~\ref{fig:spectrum}.}
\begin{ruledtabular}
\begin{tabular}{ l | c | c }
type & rel. noise power (arb. u.) &  rate (MHz)  \\ \hline \hline
Lorentzian & $A_L=2.16 \pm 0.03$ &  $2 \pi\gamma_e=27.1 \pm 1.8^\ast$      \\ \hline
Maxwell    & $A_M=1.04 \pm 0.02$ &  $2 \pi\delta_e=182.8 \pm 4.7~$   \\\hline
Gaussian   & $A_G=3.22 \pm 0.08$ &  $2 \pi\delta_e^\ast= 213.6 \pm11.3~$  \\
 \end{tabular}
\end{ruledtabular}
{\it $^\ast$ reduces to $2 \pi \gamma_e=$ $3.14\pm 1.1$~MHz for negligible excitation density.}
\label{tab:parameters}
\end{table}

All extracted fit parameters are summarized in Table~\ref{tab:parameters}. The consistency is demonstrated by the general conservation of SN power: $A_G=A_L+A_M$, which describes the redistribution of SN power from the Larmor precession peak at $B\ne 0$ towards the two-peak structure at $B=0$. However, the width of the precession peak at $\nu_L$ is increased by about $15 \%$ compared to the Overhauser contribution which is attributed to an electric field dependent $g$-factor~\cite{inhomB,poisson}, see Ref.~\cite{suppl}.

Interestingly, the power ratio of the homogeneous and the Overhauser contribution $A_L/A_M \approx 2$  deviates strongly from the expected $1/2$ ratio \cite{glazov_spin_2012,glazov_spin_2015} and is caused by  a finite value of the electron correlation time at a given donor: Overall, the two $B=0$ features are very well modeled after Eqs. (6) and (9) of Ref.~\cite{glazov_spin_2015} including (i) spin precession in the random hyperfine fields and (ii) the finite correlation time $\tau_c$.  The model  (red curve in Fig.~\ref{fig:spectrum}a) is fitted to the data with $\tau_c$ and $\delta_e$ being the only parameters and allows to extract the nuclear field spread $\delta_e\approx 29\pm0.3$~MHz. This corresponds to $\Delta_B\approx 4.6$~mT, which is in close agreement with other experimental data~\cite{PhysRevB.66.245204,colton_dependence_2004} as well as with the value extracted from the Maxwellian fit. The correlation time of $\tau_c \approx 32\pm 0.6$~ns is very close to the value reported in Ref.~\cite{PhysRevB.66.245204} for a comparable electron density. The Lorentizan fit of the zero-frequency peak gives a similar value of $\tau_c = 1/(2\pi \gamma_e) \approx 37$~ns. Nevertheless, all correlation times are mainly limited by optical excitation as discussed in the next paragraph.

  In order to gain further insight into the electron SN, the dependence of the noise power of the zero-frequency component on the photo-generated excitation density $n_{\rm ex}$ is measured by reducing the bandwidth of the used detector. This allows to accumulate SN at very low optical powers. The excitation density is calculated via  $n_{\rm ex}=P\tau/\hbar \omega  \times (1-\exp[-\alpha(\hbar \omega)d])$ from the experimental parameters ($\hbar \omega$ is the laser energy, $d$ is the layer thickness, $P$ is the optical power incident on the sample) assuming a peak absorption coefficient of the donor electron ensemble $\alpha=4000~\text{cm}^{-1} $ and a radiative lifetime of $\tau=1~$ns~\cite{finkman_oscillator_1986}. The measured data are fitted by the Lorentzian function $S_L(\nu)$ and the resulting dependence $\gamma_e(n_{\rm ex})$ is depicted in Fig~\ref{fig:spectrum}b over more than three orders of magnitude. The extrapolation towards $n_{\rm ex}=0$ yields a value of $2 \pi \gamma_e=3.14 \pm 1.1~$MHz corresponding to a correlation/spin relaxation time of about 320~ns. This time is comparable with the nuclear spin precession time in the Knight-field of the electron.~\cite{suppl}

\begin{figure}[tb]
  \includegraphics[width=0.99 \columnwidth]{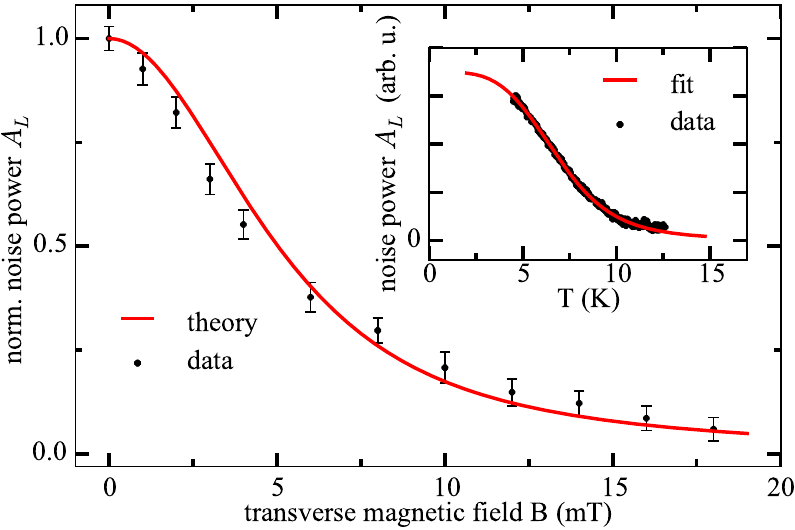}
  \caption{(Color online.)
   SN power of the homogeneous $B=0$ contribution as function of the transverse magnetic field $B$ at $T=4.2~$K and $P=270~$nW. The red line is calculated with the model in Ref. \cite{glazov_spin_2015} with the same parameters as in Fig.~\ref{fig:spectrum}a (note, that this theoretical curve is weakly sensitive to the particular value of $\tau_c$). Inset shows the dependence of $A_L$ on the cryostat temperature measured at $B=0~$mT, $P=2~\mu$W, and a detuning $\Delta=200~\mu$eV. The red line is a fit based on Blakemore's equation~\cite{blakemore_semiconductor_1987}.}
\label{fig:dBKNGIHT}
\end{figure}

 The measured noise power of the zero-frequency SN contribution as a function of the transverse magnetic field $B$ is plotted in Fig.~\ref{fig:dBKNGIHT} (black dots) showing clearly the expected reduction of the zero frequency peak.~\cite{glazov_spin_2012} The red line is calculated after Eq. (13) of Ref.~\cite{glazov_spin_2012} with the same parameters as used to fit the SN spectrum in Fig.~\ref{fig:spectrum} and shows an excellent agreement. The inset of Fig.~\ref{fig:dBKNGIHT} depicts the SN power of the zero-frequency contribution as a function of the cryostat temperature. The experimentally observed SN (black dots) reduces drastically with increasing temperature due to thermal ionization of the donors. The red line is a fit according to Blakemore's equation with the two free parameters being the doping density and a temperature offset between the sensor at the heat exchanger of the cryostat and the laser spot \cite{blakemore_semiconductor_1987}. The extracted offset is $\Delta T = 2$~K, being typical for our cryostat configuration. However, the extracted doping density of $1.5 \times 10^{12}$~cm$^{-3}$ is much lower than the nominal doping density. The origin of this discrepancy is not fully understood but could be related with unintentional $p$-type co-doping (compensation) and donor depletion due to surface charges.

\begin{figure}[htb]
 \includegraphics[width=0.99 \columnwidth]{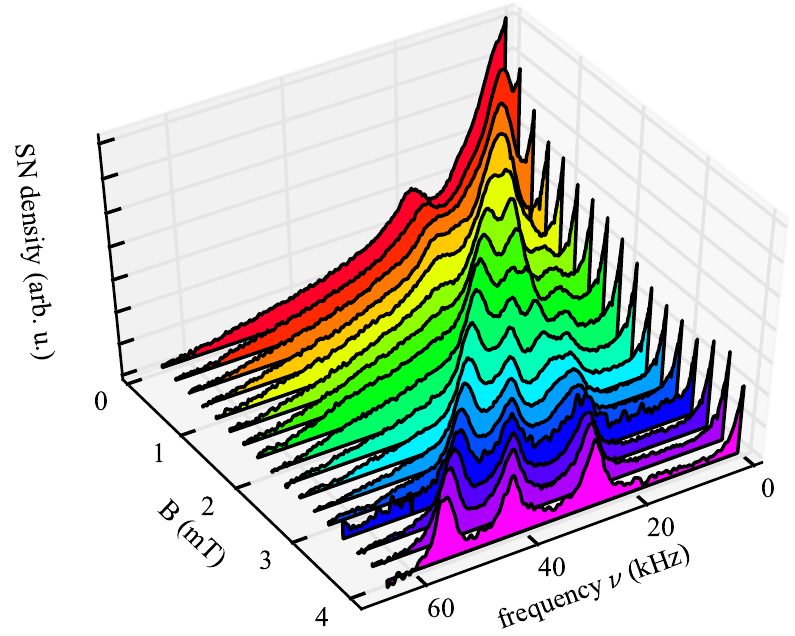}
  \caption{(Color online.)
  Nuclear SN spectra as a function of the transverse magnetic field at $T=4.2$~K and $P=1$~$\mu$W.  The homogeneous contribution (blue line in Fig~\ref{fig:spectrum}c) was subtracted.  See text for details.}
\label{fig:dB}
\end{figure}

Now we focus on the very low frequency range $1$~kHz$\ldots 70$~kHz. Here, the SN spectrum reveals a clearly resolved fine structure shown in Fig.~\ref{fig:spectrum}c acquired at a transverse field $B=3.75$~mT, where three additional, very narrow spin noise peaks at finite frequencies are resolved. These peaks shift linearly for $B\gtrsim 1$~mT, see Fig.~\ref{fig:dB}, and their origin is identified by the corresponding magnetic moments as the host lattice isotopes $ ^{75}$As, $^{69}$Ga, and $^{71}$Ga. \cite{suppl} Interestingly, the relative magnitudes of the nuclear SN do not scale with the different abundance and coupling strengths \cite{Paget.PRB.1977} of the isotopes. The origin is unclear so far. By fitting the corresponding contributions by Lorentzians~ (red line in Fig.~\ref{fig:spectrum}c), we extract a ratio $\xi = 1.2 \times 10^{-3}$ of the nuclear SN power for all host lattice isotopes to the zero-frequency electron SN power contribution $A_L$. \cite{suppl}

The observation of nuclear SN in the FR noise spectrum is, at first glance, very surprising, since lattice nuclear spins do not couple directly with light. However, nuclear spin fluctuations affect via hyperfine interaction the electron spin degrees of freedom and manifest them-self in the optical response~\cite{artemova85,PhysRevLett.111.087603}. Particularly, for bulk semiconductors with donor-bound electrons there are two contributions to nuclei-induced Faraday rotation: (i) the Overhauser field induced splitting of the $D^0X$ transition line, which is temperature independent, and (ii) a state-filling effect caused by the electron spin polarization in the Overhauser field, which depends on temperature~\cite{PhysRevLett.111.087603}. The straightforward calculation~\cite{suppl} shows that the nuclear fluctuation-induced splitting dominates the SN for quasi-resonant detection in an inhomogeneously broadened transition and that the ratio $\xi$ of the nuclei and electron spin noise powers is given by $\xi \sim (\Delta_N /\Gamma)^2$. Here, $\Delta_N\sim 0.1 \ldots 0.3~\mu$eV is the nuclear spin noise induced energetic fluctuation of the $D^0X$ line and $\Gamma$ is the \emph{homogeneous} width of the $D^0X$ resonance. Taking $\Gamma=8~\mu$eV~\cite{PhysRevB.54.4702} we estimate $\xi \sim 10^{-3} \ldots10^{-4}$~\cite{suppl} which is in rather good agreement with the experimentally observed value. This ratio is temperature independent in the studied range between $T=3.2$~K and $T=7$~K.

The detection of nuclear spontaneous spin resonance by spin noise spectroscopy provides the novel method that enables measurements of the nuclear spin dynamics without exciting a nuclear spin polarization~\cite{2015arXiv150400799R}, applying strong external magnetic fields to split the nuclear spin sublevels, or using radio frequency pulses like in NMR/ODNMR experiments. The suggested technique is particularly useful to address the nuclear spin dynamics at low magnetic fields unaccessible otherwise. Particularly, at $B\lesssim 1$~mT we reveal complex behavior of nuclear spin resonance lines with strong deviations from linear-in-$B$ dependence. These deviations, as well as the appearance of the zero-frequency line in the nuclear SN spectra can be related with small quadrupolar splittings, local fields and intricate nuclear spin decoherence~\cite{bechtold_three_2014}, see Ref.~\cite{suppl} for details, and require further in-depth studies.

In summary, detailed spin noise measurements on the neutral exciton transition of nearly isolated, localized donor electrons in GaAs yield a comprehensive picture of the intricate electron and nuclear spins at thermal equilibrium including (a) the homogeneous and Overhauser SN contribution at $B=0$, (b) the influence of the correlation time on their shape and relative noise powers, (c) the temperature dependence of the ionization of a low-density electron ensemble, (d) the inhomogeneous broadening of the Overhauser contribution at finite external magnetic fields due to electron $g$-factor variations, and (e) the observation of nuclear fluctuations by optical spin noise spectroscopy and their intricate magnetic field dependence. Especially, the new nuclear SN technique gives an inimitable access to the nuclear spin dynamics at thermal equilibrium and very low external magnetic fields and promises a variety of applications, i.e., for highly sensitive spatially resolved nuclear magnetic resonance.

\begin{acknowledgments}
We acknowledge the financial support by the BMBF joint research project Q.com-Halbleiter (16KIS0109 and 16KIS00107) and the Deutsche Forschungsgemeinschaft (TRR160 and OE 177/10-1). MMG is grateful to the Dynasty Foundation, RFBR, RF President grant MD-5726.2015.2, Russian Ministry of Education and Science (Contract No. 11.G34.31.0067), and SPbSU Grant No. 11.38.277.2014.
\end{acknowledgments}


\begin{thebibliography}{48}
\bibitem [{nor()}]{normA}%
  \BibitemOpen
  \href@noop {} {}\bibinfo {howpublished} {The constants $A_M$, $A_L$ and $A_G$
  represent the half of the SN power, since SN spectra are symmetric with
  $\nu\to -\nu$ replacement.}\BibitemShut {Stop}%
\end{thebibliography}

%

\end{document}